\documentclass[useAMS,usenatbib,usegraphicx]{mn2e}

\usepackage{times}

\newcommand{\<}{\begin{eqnarray}}
\renewcommand{\>}{\end{eqnarray}}
\renewcommand{\bar}{\overline}

\renewcommand{\hat}{\widehat}
\newcommand{\age}{\mathrm{age}}
\newcommand{\SFR}{\mathrm{SFR}}
\newcommand{\Msun}{\rmn{M}_\odot}
\newcommand{\Myr}{\mathrm{Myr}}
\newcommand{\yr}{\mathrm{yr}}
\newcommand{\dex}{\mathrm{dex}}
\newcommand{\mx}{\mathrm{max}}
\newcommand{\MAX}{\mathrm{MAX}}

\newcommand{\MIN}{\mathrm{MIN}}

\newcommand{\tot}{\mathrm{tot}}

\newcommand{\cl}{\mathrm{cl}}

\title[Star formation histories of galaxies]{
A new method to derive star formation histories of galaxies from their star cluster distributions
}
\author[Th. Maschberger and P. Kroupa]
{Thomas Maschberger\thanks{e-mail: tmasch@astro.uni-bonn.de} and Pavel Kroupa\thanks{e-mail: pavel@astro.uni-bonn.de}\\
Argelander-Institut f\"ur Astronomie, Auf dem H\"ugel 71, Bonn\\
Rhine Stellar Dynamics Network (RSDN)}
\date{2007}

\pagerange{\pageref{firstpage}--\pageref{lastpage}} \pubyear{2007}

\begin{document}
\label{firstpage}

\maketitle

\begin{abstract}
Star formation happens in a clustered way which is  why the star cluster population of a particular galaxy is closely related to the star formation history  of this galaxy.
From the probabilistic nature of a mass function follows that the mass of the most-massive cluster of a complete population, $M_\mx$,  has a distribution with the total mass of the population as a parameter.
The total mass of the population is connected to the star formation rate (SFR) by the length of a formation epoch.

Since due to evolutionary effects only massive star clusters are observable up to high ages it is convenient to use this $M_\mx (\SFR)$ relation for the reconstruction of a star formation history.
The age-distribution of the most-massive clusters can therefore be used to constrain the star formation history of a galaxy.
The method, including an assessment of the inherent uncertainties, is introduced with this contribution, while following papers will apply this method to a number of galaxies.
\end{abstract}


\section{Introduction}
During the last few years it has been recognised that most and probably all stars form in embedded clusters \citep{lada+lada2003}.
The observational work, notably by \citet{larsen2002},  has established that star clusters ranging from the oldest globular clusters to the youngest low-mass objects have to be regarded as a continuous distribution by mass --- globular clusters are not fundamentally different from open clusters but merely the upper mass end of the distribution.
This has been shown explicitly by \citet{kroupa+boily2002} on the example of the Milky Way population II spheroid.
Today star clusters must be viewed as the {\it ``fundamental building blocks''}\  of galaxies because they also determine the morphological appearance of whole galaxies through the physics of their formation \citep{kroupa2005}.

Star formation is therefore  closely connected to the star cluster distribution in a galaxy  and thus it is expected that the star formation history of a galaxy leaves its imprint on the star cluster distribution.
Since star clusters can live for a long time the star cluster distribution of a galaxy can be compared to a diary:
Events like interactions of galaxies  lead to an enhanced star formation activity.
From this results a larger number of clusters being formed during the interaction time.
This qualitative statement is well known and in this work we present a quantitative method to derive the star formation history (SFH) of a galaxy directly from its star cluster content.

Until now detailed SFHs can be determined only for galaxies which are at a distance allowing individual stars to be resolved, i.e. within the local group up to $\sim$ 1 Mpc away.
The colour-magnitude diagram (CMD) which is thusly obtained  then allows the construction of a SFH using theoretical isochrones.

In distant galaxies, however, star clusters  appear as compact sources whereas the individual stars give a homogeneous distribution over the area.
Modern instruments like the Hubble space telescope make it possible to obtain a cluster age- and mass-distribution for galaxies beyond the distance where individual stars can be resolved, e.g. for M51 
(\citealp{bastian-etal2005}; cluster formation rate: \citealp{gieles-etal2005a}) 
and M101 \citep{bianchi-etal2005}.
These galaxies lie at a distance of about 7 Mpc (M51: \citealp{takats+vinko2006}; M101: \citealp{kelson-etal1996}) which demonstrates the potential of our new method.

A first approach to derive the star formation history from a cluster age- and mass-distribution could simply be to use all observed clusters in given time intervals.
But this would lead to wrong results, clusters evolve and thus there are fewer clusters at higher ages.
However, massive clusters evolve slowly, clusters with masses  $\ga 10^5\ \Msun$ have lifetimes comparable to a Hubble time \citep{baumgardt+makino2003}.
\citet*{weidner-etal2004} found a relation which establishes the connection between massive clusters and the galaxy-wide star formation rate:
During a {\it formation epoch} a {\it complete population} of star clusters is formed, for which the mass of the most massive cluster, $M_\mx$, depends on the star formation rate (the $M_\mx (\SFR)$ relation).
Consequently, the star formation history of a galaxy can be regarded as a sequence of such formation epochs, in which the most massive cluster carries the information about the star formation rate.

Since \citet{weidner-etal2004} used a deterministic law for the $M_\mx (\SFR)$ relation, we briefly re-analyse their argumentation, allowing a distribution of the most massive cluster rather than a fixed value for a given SFR.
Then we present the new method and conclude with testing it for some typical cases.

\section{Complete Populations and the $M_\mx (\SFR)$-Relation}
\subsection{Complete Populations}
The distribution of star clusters in a star cluster population is characterised by the shape of the distribution function (shown to be a power law, e.g. \citet{weidner+kroupa2006} and references therein), the mass of the population, $M_\tot$, and the mass limits, $M_\MIN$ and $M_\MAX$.
The cluster mass function can be written as
\< \xi_\cl (M) = \frac{d N}{d M} &=& k M^{-\beta}, \>
where the normalisation constant $k$ is determined by $M_\tot$, $M_\MIN$ and $M_\MAX$.

In this context {\it complete population} denotes a statistically meaningful representation of the embedded cluster mass function.
The underlying distribution function is defined on a mass interval appropriate to the mass of the population.
This is a very general concept which is applicable to other cases where objects obey a distribution function, e.g. stars in a star cluster.

The lower mass limit, $M_\MIN$, is given by the physical minimum mass of a star cluster and is independent of the mass of the population.
Until now, the physical minimal mass of a star cluster is not well known, here it is assumed to be 5 $\Msun$ corresponding to groups of about a dozen stars such as those forming in Taurus-Auriga \citep{briceno-etal2002}.

For the upper mass limit, $M_\MAX$, two cases have to be distinguished:
If a complete population has a mass which is larger than the limiting physical maximum cluster mass, then the distribution function is defined up to the physical maximum mass, a hitherto not well understood quantity.
Star clusters with masses larger than about $10^6 \Msun$ show complex stellar populations and are probably distinct from the ``normal'' star cluster content of a galaxy \citep{weidner-etal2004}.
Such arguments based on the structural properties of clusters would imply the same  physical maximum mass in all galaxies.
\citet{gieles-etal2006} derived maximal cluster masses from the cluster luminosity function, obtaining masses between $0.5-2.5 \times 10^6 \Msun$ for NGC 6946, M51 and the ``Antennae'' (NGC 4038/39).
Thus, while arguments exist for a limiting maximum star-cluster mass below which stellar populations are simple (mono-metallic and -age), \citet*{mieske-etal2002} and \citet{martini+ho2004} show that ultra-compact dwarf galaxies ($M \ga 10^6 \Msun$) may be an extension of the ``cluster'' formation process to large masses.
Consequently, we do not limit the  ``cluster'' masses but allow these to formally reach $M_\MAX = 10^9 \Msun$ for sufficiently high SFRs.

The case that $M_\tot$ is smaller than the physical maximal mass implies that no star clusters more massive than $M_\tot$ can exist in this population.
Therefore $M_\tot$ is the upper mass limit of the cluster mass function.
This includes the case that a population can consist only of one cluster with the mass $M_\tot$, but this is very improbable.

Thus by using the total mass as the normalisation criterion,
\< M_\tot &=& \int_{M_\MIN}^{M_\MAX} M \xi_\cl (M) d M,\>
 the normalisation constant becomes
\< k &=& \frac{M_\tot (2- \beta)}{M_\MAX^{2-\beta} - M_\MIN^{2-\beta}}. \>
The total number of clusters in a population, $N_\tot$, follows from $N_\tot (M_\tot) = \int_{M_\MIN}^{M_\MAX} \xi_\cl (M) d M.$

\citet{weidner-etal2004} argued that a complete population of star clusters is not made up by all clusters ever formed in a galaxy, but by a subset of clusters formed during a {\it formation epoch}.
Assuming that all stars form in star clusters the total mass of a complete population is then given by the product of the SFR and the length of the formation epoch,
\< M_\tot = \SFR \times \delta t. \label{mtotdeltat}\>
Thus, given a certain (short) $\delta t$, the total mass and thereby implicitly the upper mass limit of the distribution function depend on the current SFR.
With this description massive clusters can only form if there is much star forming activity, while quiescent phases only produce  low-mass clusters.

\subsection{\label{mmxsfr}The $M_\mx (\SFR)$ relation}

\begin{figure}
\includegraphics[width=8cm]{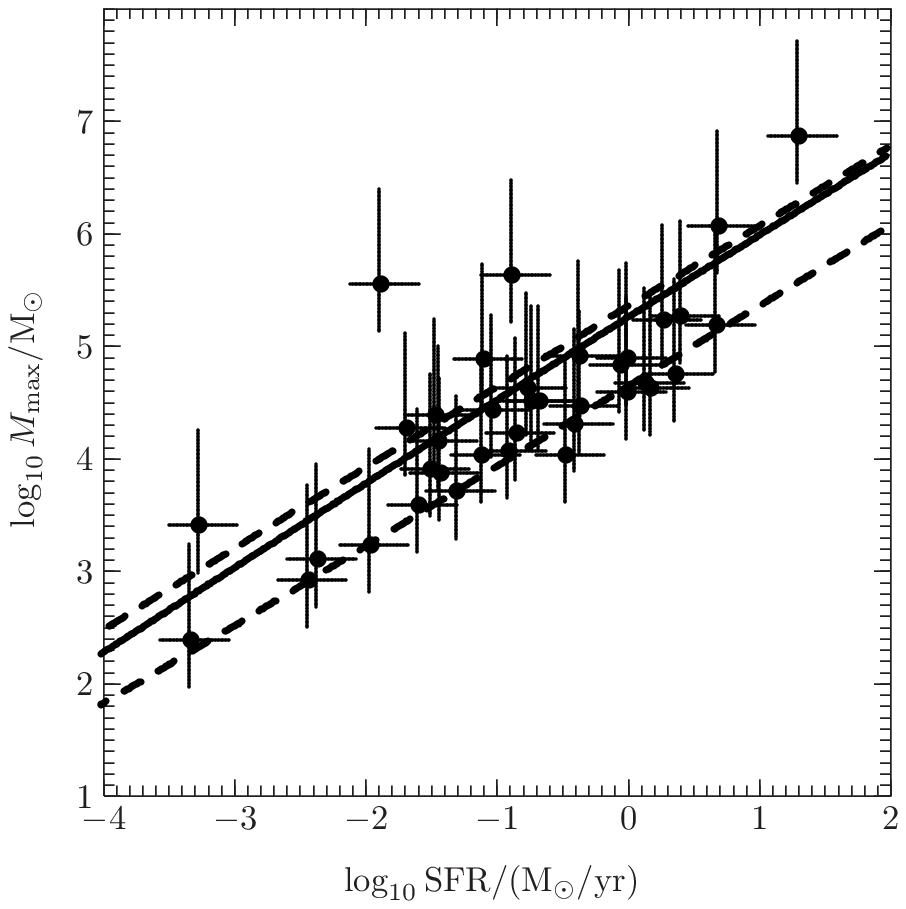}
\caption{
Masses of the brightest clusters vs. presen-day SFR in galaxies, data from \citet{larsen2002} (absolute magnitudes converted to masses by \citealp{weidner-etal2004}).
The lines shown with the data are based on our statistical point of view, presented in Sec.~\ref{mmxsfr}.
The parameters $\delta t$ and $\beta$  are chosen to fit to the data with $\delta t = 10 \Myr$,  $\beta = 2.4$, and $M_\MIN= 5 \Msun$.
The solid line is the expectation value for the distribution of the most massive cluster (eq.~\ref{barmmax}) in dependence of the total mass of the population.
The dashed lines are the borders of the region in which 2/3rd of all most massive clusters are expected, calculated using eq.~\ref{mmaxpdf}.
}
\label{figmmaxsfr}
\end{figure}

Observations give evidence that the brightness of the brightest cluster in a galaxy depends on the present-day SFR in the galaxy \citep{larsen2002}.
Since the brightest cluster in a galaxy is usually young it can be interpreted as the most massive cluster of the current formation epoch.
\citet{weidner-etal2004} converted the luminosities to masses, as shown in Fig. 1.
The data show a large scatter which in our interpretation results from the distribution of the most massive cluster.
Since the distribution of $M_\mx$ for a given $M_\tot$ is known the data can be used to determine the length of the formation epoch.

\begin{figure}
\includegraphics[width=8cm]{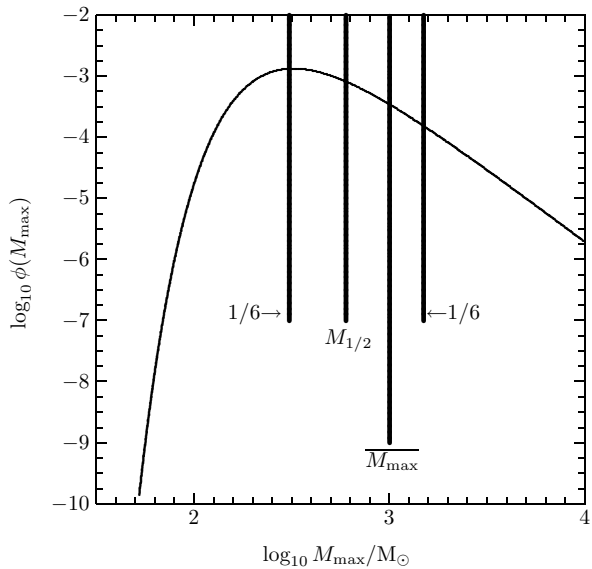}
\caption{The distribution of $M_\mx$ for a population with $M_\tot= 10^4 \Msun$, $\beta=2.4$, and $M_\MIN = 5 \Msun$. The number of $M_\mx$ expected to lie below and above the average $\bar{M_\mx}$ is different due to the asymmetry. The median, $M_{1/2}$, has a different value than the average. 2/3rd of all $M_\mx$ are expected in the region delimited by lines marked with $1/6 \rightarrow $ and $\leftarrow 1/6 $.
}
\label{figmmaxpdf}
\end{figure}

In our description of star cluster populations the upper mass limit, $M_\MAX$ and the most massive cluster, $M_\mx$, are not identical.
For an ensemble of populations with the same total mass, $M_\mx$ has a distribution parametrised by $M_\tot$ and $\beta$ \citep[cf.][]{oey+clarke2005}.
This distribution can be written as a probability density,
\<  \phi (M_\mx) &=& \left( \frac{1}{N_\tot} \int_{M_\MIN}^{M_\mx} \xi_\cl (M) d M \right)^{N_\tot-1}  \xi_\cl (M_\mx),   \label{mmaxpdf}\>
where $N_\tot$ and the normalisation of $\xi_\cl$ depend on $M_\tot$.
This is different to the ansatz of \citet{weidner-etal2004} where $M_\mx$ was assumed to be identical for all populations with the same $M_\tot$, i.e. not distributed.
The distribution of $M_\mx$ is asymmetric because of the asymmetric cluster mass function and is characterised by the average mass of the most massive cluster, $\bar{M_\mx}$, given by
\< \bar{M_\mx}  &=& \int_{M_\MIN}^{M_\MAX} M_\mx' \phi ( M_\mx') d M_\mx'. \label{barmmax} \> 
Figure~\ref{figmmaxpdf} shows the distribution of $M_\mx$ and the location of $\bar{M_\mx}$.
Due to the asymmetry the median, $M_{1/2}$, does not have the same location as the average, $\bar{M_\mx}$, but lies below it.
Therefore it is expected that in an ensemble more $M_\mx$ lie below $\bar{M_\mx}$ than above.

The observations shown in Fig.~\ref{figmmaxsfr} are an ensemble of most massive clusters of populations with different total masses.
Since $ \phi(M_\mx) $ is parametrised by $M_\tot$ it is possible to derive the average $\bar{M_\mx}$ in dependence of $M_\tot$ and via eq.~\ref{mtotdeltat} also in dependence of the SFR.
This analytical curve could be compared to a line derived from observations e.g. by least squares fitting.
The least squares fitting procedure would give a line that leads through the region where most of the $M_\mx$ are, but this does not match $\bar{M_\mx}$ due to the asymmetry of $\phi (M_\mx)$.
Most of the $M_\mx$ lie below $\bar{M_\mx}$.

To constrain $\delta t$ from the observations it is more convenient to use the region where a certain fraction of the data is expected.
The location of this region -- at high or low values of $M_\mx$ -- depends on the duration of the formation epoch:
A long formation epoch pushes the region towards high masses,  a short formation epoch to lower.
A larger $\beta$ steepens the relation.
The best fitting region where 2/3rd of the most massive clusters are expected is shown in Fig.~\ref{figmmaxsfr}.
As found by \citet{weidner-etal2004}  the values of $\delta t$ = 10 Myr and $\beta = 2.4$ provide a good fit to the data.
Complete populations of star clusters form in formation epochs lasting for 10 Myr which is comparable to the time-scale of the emergence of cluster populations from spiral arms (\citealp*{egusa-etal2004}; \citealp{bonnell-etal2006}).

\section{\label{sfhmethod}A method to derive star formation histories using star clusters}
\subsection{Concept and restrictions}
The analysis of the $M_\mx (\SFR)$ relation suggests that the SFH of a galaxy can be interpreted as a sequence of formation epochs lasting for $\approx$ 10 Myr.
A star cluster population with a total mass determined by the current SFR emerges during each formation epoch.
Therefore it should be possible to infer the SFR from the properties of this population, but for the largest part of the lifetime of a galaxy only a small fraction of coeval clusters is observable making the determination of the population mass difficult.

On this account we propose a different approach using only the most massive clusters of each formation epoch.
The mass of the most massive cluster also depends on the current SFR, but the distribution of them does not allow the  derivation of the SFR from only {\it one} most massive cluster.
If it is assumed that the galaxy-wide SFR changes significantly only on a time-scale that includes a number of formation epochs, then the clusters of this set of formation epochs can be seen as an ensemble of identical cluster populations.
For the ensemble average of the most massive clusters the mass of the population can be calculated using the $M_\mx (\SFR)$ relation:
The probability density $\phi (M_\mx)$ is parametrised by $M_\tot=\SFR \times \delta t$ and $\beta$ and consequently also $\bar{M_\mx}$, written symbolically as
\< \bar{M_\mx} &=& f(\SFR), \>
where $f(\SFR)$ is given by the integral of eq.~\ref{barmmax}.
Thus the SFR corresponding to this ensemble follows by inversion of the previous equation,
\< \SFR &=& f^{-1} (\bar{M_\mx}). \>
The inversion of the integral ($f$) is done numerically.

This gives the general idea to reconstruct SFHs: The lifetime of a galaxy is divided into time windows containing a number of formation epochs.
For each of the windows the average mass  of the most massive clusters is calculated and from this mass the underlying SFR is derived. 

The age determinations of available cluster data for galaxies, as e.g. the Large Magellanic Cloud \citep{degrijs+anders2006}, usually have uncertainties of $\approx$ 0.4 dex.
If the observational situation is optimal, i.e. observations in the most suitable filters could be made, then the age uncertainty can be much smaller.
\citet{degrijs-etal2005} obtained $\Delta \log (\age/\yr) \la 0.15$ ``in the majority of cases''  for conditions as for NGC 3310. 
Since it does not make any sense to try to detect variations of the SFR on time-scales shorter than the age uncertainties of the data, the averaging window has to be chosen to have the same length (or longer) than the available age uncertainty.
Because the error is constant in logarithm the length of the averaging window depends on the age.
We chose a length of 0.5 dex for the averaging window.
A sequence of neighbouring, independent averaging windows leads to a SFH determined at discrete points in time over the lifetime of a galaxy.
For a continuous SFH the averaging window is moved in 10 Myr-steps.
At old ages the length of the averaging window is reduced to ensure that the oldest formation epoch used in the averaging window contains a cluster.
An averaging window filled only halfway would lead to a systematically underestimated SFR.

As the averaging window increases with age, the minimum duration of resolvable events in the SFH also increases.
This is an inherent restriction of our proposed method.
Short bursts of star formation which happened at large ages cannot be resolved.

\subsection{Corrections for dynamical cluster evolution}
The cluster ages and masses for large samples of clusters are usually determined by fitting models to observed spectral energy distributions.
The age and the ``initial'' \ mass of a cluster are parameters for the fitting routine. In all cases the cluster mass is determined by scaling a model ``initial'' \ mass.
Since the applied models usually only consider mass loss due to stellar evolution and not due to dynamical evolution, the fitted mass of a cluster does not correspond to the initial cluster mass for which the $M_\mx (\SFR)$ relation is valid.

Star clusters are no isolated static objects: the gravitational force keeps the stars in constant motion with respect to the centre of mass, which itself moves on an orbit through the host galaxy.
Stars can evaporate from the region dominated by the cluster potential and  leave their star cluster.
This mass loss of the cluster due to dynamical evolution depends on the eccentricity of the orbit and the distance to the galactic centre.
Furthermore, it depends on the mass of the host galaxy which determines the strength of the tidal field.

Since these parameters are mostly not available, the analytic model from \cite{lamers-etal2005a} for the average mass loss suffered by a cluster in a galaxy is used.
\citet{lamers-etal2005a} find a good agreement of their model with the results of \citet{baumgardt+makino2003}, who give a formula derived from $N$-body experiments.

For the statistical model adopted here it is assumed that cluster disruption depends only on the initial mass \citep{boutloukos+lamers2003}.
In this case, the initial mass can be calculated from the observed mass and age using
\< M_i (t) &=& \left( \left( \frac{M}{\Msun} \right)^{\gamma} + \frac{\gamma t}{t_0} \right)^{\frac{1}{\gamma}}, \label{evolution} \>
where $\gamma=0.62$ is identical for all galaxies.
$t_0$ describes the tidal field and can be determined from the dissolution time of a $10^4 \Msun$ cluster,
\< t_0 &=& \left(\frac{t_4}{660}\right)^{\frac{1}{0.967}}. \>
The parameter $t_4$ has been determined for a number of galaxies \citep*[cf.][]{boutloukos+lamers2003,lamers-etal2005}.

\subsection{Upper and lower limit for the SFH}
The observed cluster content of a galaxy usually does not provide a cluster for every formation epoch.
The number of observed clusters older than a few Gyr is much smaller than the number of formation epochs. 
\citet{degrijs+anders2006} found only $\approx$ 10 clusters older than 4 Gyr in the Large Magellanic Cloud,  as similarly \citet{bastian-etal2005} for M51.
This originates from a SFR which was so low that no clusters were formed being massive enough to be visible today.
The brightness limit of the observations is therefore an upper limit for the brightness of the actually formed most massive cluster.
With a cluster evolution model the brightness limit at a given time can be converted to an initial mass which then is the upper mass limit for the most massive cluster.
The GALEV models \citep{schulz-etal2002} give the luminosity evolution $\mathcal{M} (t)$ of a  ``simple stellar population''  \ (i.e. single burst, single metallicity) with a mass of $1.6 \times 10^9 \Msun$ for different photometric bands.
By scaling $\mathcal{M} (t)$ to the limiting magnitude of the observation, $\mathcal{M}_\mathrm{lim}$, with the band chosen according to the bands used in the cluster-mass fitting, the limiting mass can be derived \citep[cf. ][]{hunter-etal2003}:
\< M_\mathrm{lim} (t) &=&1.6 \times 10^{9 + 0.4 (\mathcal{M} (t) - \mathcal{M}_\mathrm{lim})} \Msun. \>
The GALEV models do not take dynamical evolution  into account, therefore $M_\mathrm{lim}$ has also to be corrected for dynamical evolution as described in the previous section ($M = M_\mathrm{lim}$ in eq.~\ref{evolution}).

Now the SFH can be derived with $M_\mathrm{lim}$ as the most massive cluster in those formation epochs that do not contain any clusters, leading to the upper estimate of the SFH.
Since $M_\mathrm{lim}$ increases for older ages due to the internal evolution of clusters, the derived SFR also increases, which does not necessarily reflect the underlying SFH.

The lower limit of the SFH is given by using $M_\mx=0$ in the epochs containing no cluster, corresponding to the assumption that during these epochs no star formation took place at all.

\subsection{Self-consistency checks}
In the next section tests to analyse the systematic errors of our method are made using modelled SFHs.
For an observed nearby galaxy the obtained results can be compared to independently determined quantities.
The SFH derived using star clusters should be similar to the one obtained using colour-magnitude diagrams.
This comparison can be difficult if the used regions of the galaxy differ.

The total mass of stars in a galaxy can be calculated since with our method  a SFH for an entire galaxy is derived.
For each 10-Myr epoch a SFR has been determined, from which the mass of the formed stars can be calculated.
The sum over all epochs gives the mass of the stellar content of the galaxy.
When deriving this mass, stellar evolution is not taken into account, i.e. all stars that ever formed are counted regardless of whether they still exist or not.
This can be compared to independent determinations of the stellar content of a galaxy.

Furthermore, the cluster formation rate should reflect the structures found in the SFH.
For each most massive cluster an appropriate total number of clusters should exist.
This comparison of the cluster formation rate (i.e. number of clusters per time) and the SFH can be done e.g. by using a cluster formation rate also derived using a moving time window.

\section{Modelled star formation histories}
To test our method synthetic star cluster populations were generated for different SFHs.
The aim is to verify if an input SFH can be re-extracted and to study the effects of the averaging.
For this purpose we first consider the simplest case with optimal data, i.e. a constant SFR and no measurement uncertainties for the age.
Models with a varying SFR that include the age uncertainties show the capacity of our method.
For clarity and to focus on the effects of the averaging and age uncertainties, cluster evolution is only considered in the last model (Fig.~\ref{sfhlmc}). 

The general procedure of our models is as follows:
In time steps of $10\ \Myr$ complete cluster populations are generated with a total mass given by eq.~\ref{mtotdeltat} and the mass limits $M_\MIN = 5 \Msun$ and $M_\MAX = \min ( M_\tot, M_{\MAX, \mathrm{phys}})$ with $M_{\MAX,\mathrm{phys}} = 10^9 \Msun$. 
Then an age uncertainty is assigned to each cluster with age $\tau$, drawn randomly from a Gaussian in logarithmic age, $\mathcal{N} (\log_{10} \tau'; \mu=\log_{10} \tau,\sigma_\tau)$.
The values chosen for the variance $\sigma_\tau$ are 0.15 dex and 0.35 dex, corresponding to the typical uncertainty range of the SED fitting method \citep{degrijs-etal2005}.
Uncertainties larger than 0.5 were rejected and generated again until they are smaller than 0.5.
The procedure for synthetic cluster evolution is described in the Section of the respective model.
For each $10\ \Myr$ interval only the most massive cluster is shown in Figs.~\ref{figsfrconst}--\ref{figsfrbump}.

\subsection{\label{secsfrconst} Constant SFR}

\begin{figure}
\includegraphics[width=8cm]{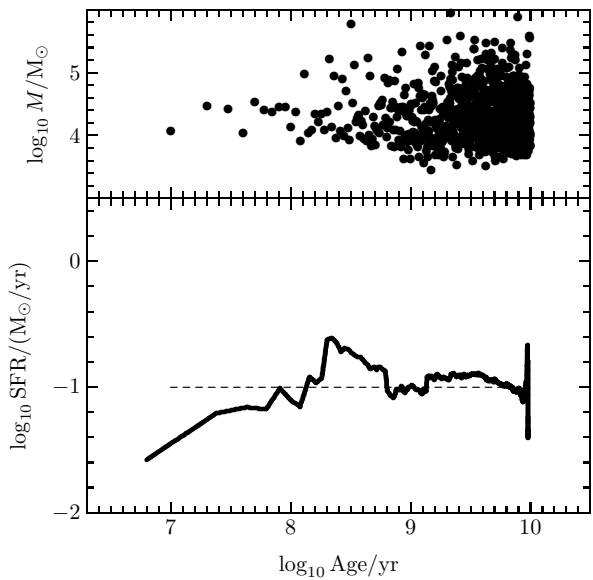}
\caption{ 
Test of the new method for a modelled constant input SFR (dashed line in the lower part). In the upper part the most massive clusters of each 10-Myr formation epoch are shown (dots). The solid line in the lower part is the reconstructed SFH using the method described in the text. 
}
\label{figsfrconst}
\end{figure}

Figure~\ref{figsfrconst} shows the distribution of the most massive clusters for a constant input SFR (dashed line in the lower part) not including age uncertainties.
It appears as if the masses of the most massive cluster increase with age, which is a  size-of-sample effect.
Due to the logarithmic axis 90 \% of the clusters are in the age range  $9 \le \log_{10} \tau \le 10$, therefore the probability to get a very massive cluster by chance is higher in this range.
Similarly the probability to sample clusters in the mass range below the average mass of the most massive cluster ($\bar{M_\mx}$, eq.~\ref{barmmax}) increases.
This leads to the wedge-like shape of the distribution of the most massive clusters.

The derived  SFH (solid line) only shows a large variation at young ages, where the large averaging window of 0.5 dex in log age contains only a few $10\ \Myr$ formation epochs.
Therefore the reconstructed SFH is calculated from too few clusters.
For ages older than $\approx 100\ \Myr$ the reconstructed SFH agrees within the expected statistical variations (as discussed in Section~\ref{statscat}) with the input SFH.
The features in the reconstructed SFH at an age of 10 Gyr are artefacts due to the shrinking averaging window.

\subsection{\label{secsfrvar} Varying SFR}

\begin{figure}
\includegraphics[width=8cm]{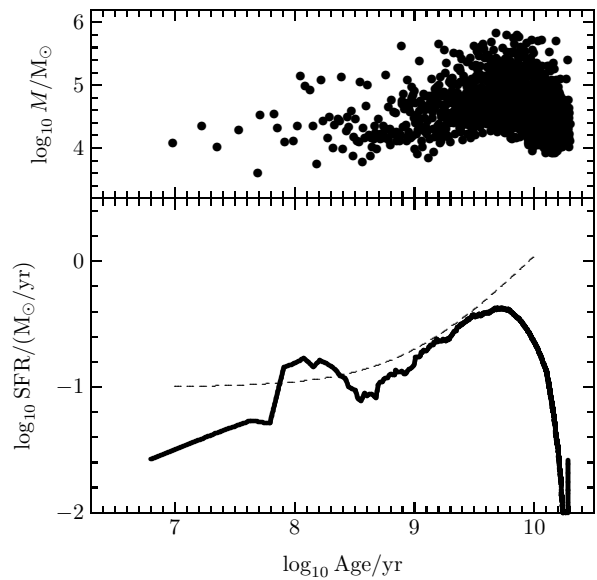}
\caption{Same as Fig.~\ref{figsfrconst} but for a slowly linearly decreasing input SFH. 
Here an age uncertainty with $\sigma_\tau = 0.15\ \dex$  was included, leading to the decline at high ages as discussed in the text.
}
\label{figsfrlin}
\end{figure}

Figure~\ref{figsfrlin} shows the SFH obtained from a linearly decreasing SFR.
In this model an age uncertainty with $\sigma_\tau = 0.15 \dex$ was included as described above.
This causes a systematic deviation of the reconstructed SFH towards smaller SFRs at old ages.  
Although cluster populations were only generated up to an age of 10 Gyr the ages of individual clusters can be allocated up to maximally $\log_{10} \tau = 10 + 0.5$. 
Thus, for many old formation epochs not the actual most massive cluster corresponding to it is used but the second or third etc. most massive.
Therefore the SFR is underestimated.
Besides this effect introduced by the age uncertainties the behaviour of the reconstructed SFH is similar to the one of the previous model:
The reconstructed SFH follows the slow change of the input SFH within the same degree of deviations. 
There is a large scatter in the  SFH for ages younger than $\approx 100\ \Myr$.
Also as in the constant case the artefacts at the oldest ages ($\log_{10} \tau \ga 10.3$) are visible. 
This model allows to conclude that our method is capable of reproducing slowly changing SFHs.

\begin{figure}
\includegraphics[width=8cm]{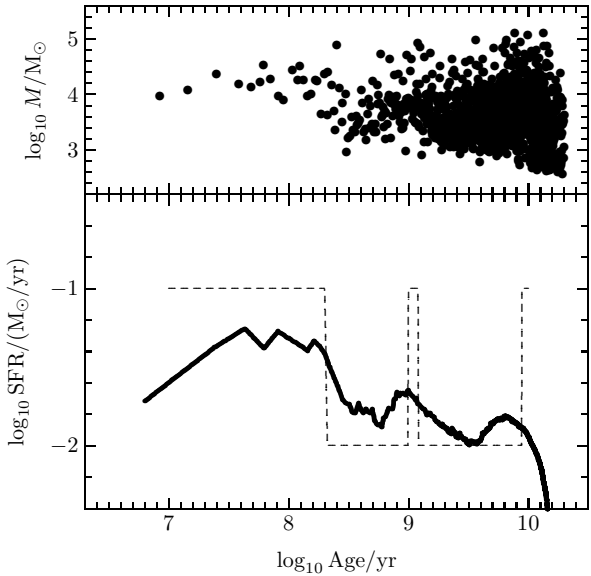}
\caption{Same as Fig.~\ref{figsfrlin} but for a SFH with three well-separated bursts.
}
\label{figsfrbump}
\end{figure}

The modelled SFH shown in Fig.~\ref{figsfrbump} has three well-separated bursts lasting from 0--200 Myr, 1 000--1 200 Myr, and 8 800--10 000 Myr.
During the bursts the SFR is increased by a factor of 10.
Again the cluster ages were generated  with an age uncertainty with $\sigma_\tau = 0.15$ dex.
Our method results in an undulating SFH with peaks roughly coinciding with the centres of the bursts.
The oldest burst is affected by the decline of the reconstructed SFH due to the way how the age uncertainties are assigned.
Thus the maximum is shifted towards younger ages.
In the reconstruction the shape of the SFH is much less pronounced than in the original.
The age uncertainties and the averaging procedure lead to a larger width of the recovered bursts.
As the most massive clusters of a burst are spread over an interval longer than the burst the amplitude of a recovered burst also decreases.

In Figures~\ref{figsfrconst} to \ref{figsfrbump} the input SFHs guide the eye to see the appropriate structure in the SFHs derived from the cluster distributions.
The distribution of the most massive clusters, the age uncertainties, and the averaging process in our method lead to a much less distinct structure in the reconstructed SFH than in the initial one.
In Section~\ref{sigcrit} we present a criterion to decide which features in a reconstructed SFH are caused by variations in the initial SFH and are not merely due to  systematic effects.
For this purpose it is necessary first to investigate the systematic effects of our method, which is the object of the next Section.

\subsection{\label{statscat}Statistical scatter in the reconstructed SFH}
The models presented above show that the input and the reconstructed SFHs differ, especially for ages younger than $\approx$ 100 Myr.
Due to the small number of clusters used for the reconstructed SFH in the respective formation epochs the scatter in the reconstructed SFH increases.
To investigate the expected scatter in the reconstructed SFHs, a sample of star cluster populations with the same SFH was created.
For each of the input SFHs, 1000 synthetic cluster populations (i.e. 1000 galaxies) were generated with different initial random seeds and the SFH was re-extracted from the clusters.
Then the sample average, $\bar{\SFR} (t) $, was calculated for each formation epoch,
\< \bar{\SFR} (t) &=& \frac{1}{1000} \sum_{i=1}^{1000} \SFR_i (t), \>
where $\SFR_i (t)$ is the  SFH of an individual cluster population $i$.
To achieve an estimate of the statistical spread we calculated the positive and nevative deviation, $\sigma_{\SFR+}$ and $\sigma_{\SFR-}$, from the average SFR at a given time,
\< \sigma_{\SFR+} (t) &=& \frac{1}{N_+ (t)} \sum_{\SFR_i (t) - \bar{\SFR} (t) > 0}  |\SFR_i (t) - \bar{\SFR} (t) | \label{sigmaplus} \\ 
 \sigma_{\SFR-} (t) &=& \frac{1}{N_-(t)} \sum_{\SFR_i (t) - \bar{\SFR} (t) < 0}  |\SFR_i (t) - \bar{\SFR} (t) |, \label{sigmaminus} \>
where $N_\pm(t)$ is the number of SFRs larger or smaller than $\bar{\SFR}(t)$ at a given time $t$.
This particular choice of an individual positive and negative average deviation will be discussed after the average SFHs.

\begin{figure}
\includegraphics[width=8cm]{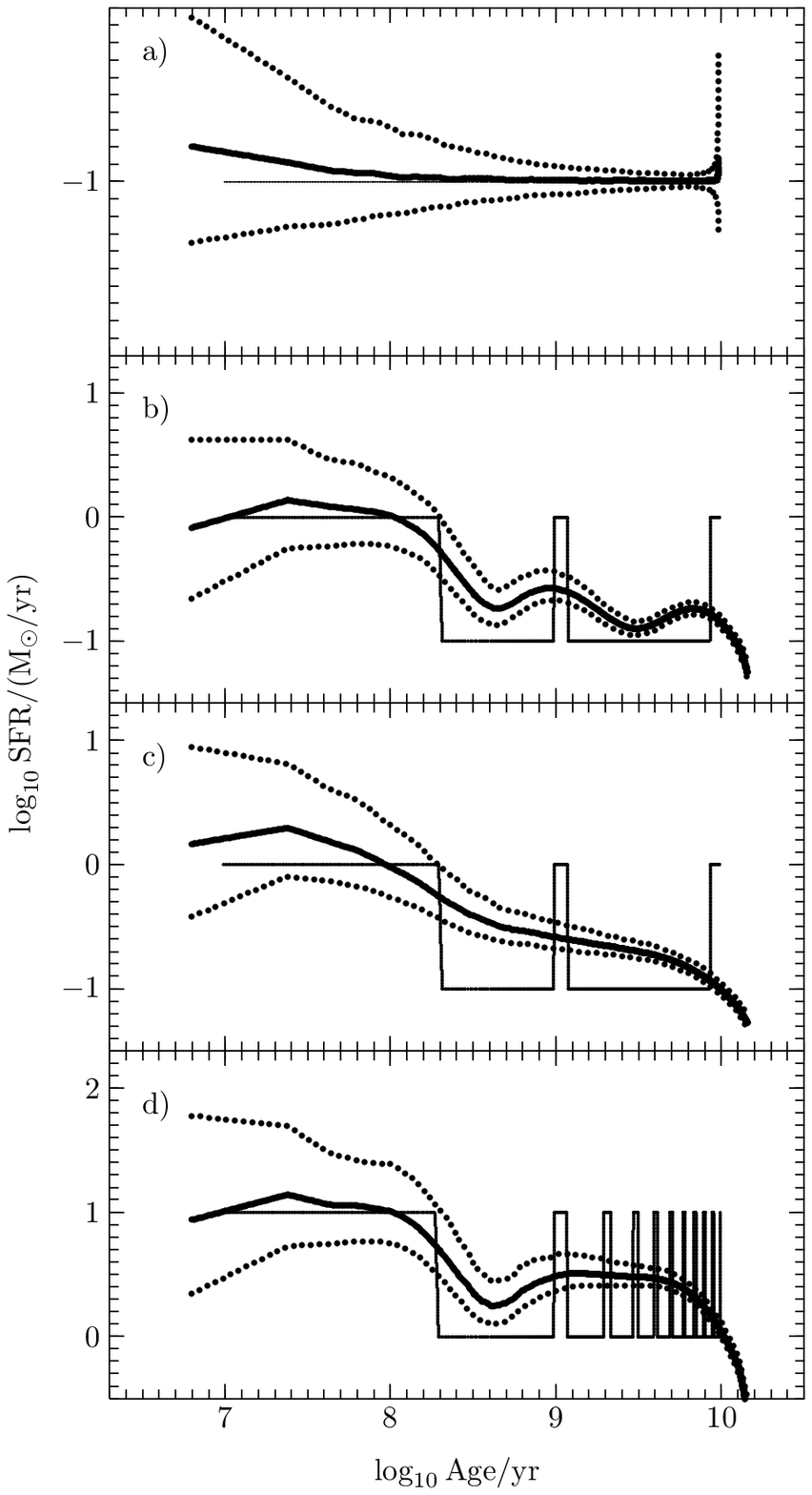}
\caption{Averaged reconstructed SFHs (thick solid lines) of a sample of 1000 synthetic cluster populations for different input SFHs (thin solid lines).  Also shown are the corresponding deviations, $\sigma_{\SFR+}$ and $\sigma_{\SFR-}$, as discussed in the text (dotted lines, eq.~\ref{sigmaplus} and \ref{sigmaminus}).  \newline
a) constant input SFH, no age error.\newline
b) three well separated bursts, age error $\sigma_\tau = 0.15$ dex.\newline
c) three well separated bursts, age error $\sigma_\tau = 0.35$ dex.\newline
d) one burst per Gyr, age error $\sigma_\tau = 0.15 $ dex.
}
\label{figsfrsigma}
\end{figure}

Figure~\ref{figsfrsigma} shows the results of the experiments.
For the experiment presented in panel a) a constant SFH without age errors was chosen.
The sample average and the input SFH agree well, differences occur only for young ages.
This is because of the small number of formation epochs used for reconstructing the SFH at these ages.
Since the averaging window contains few epochs the scatter in the reconstructed SFH increases, which is also visible in the progress of $\sigma_{\SFR\pm}$.
The average deviations decrease with age because the number of formation epochs used for the reconstructed SFH increases as a consequence of the averaging window which moves in logarithmic time.
At the oldest ages the average deviation increases again since the averaging window contains a decreasing number of formation epochs with a cluster, which is the same effect as for young ages. 
As described in Section~\ref{secsfrconst} the SFH ends at the oldest ages with an artifact.

\begin{table}
\caption{Original and reconstructed total masses of the stellar content for the different input SFHs. Output $M_\tot$ is the ensemble average of the 1000 galaxies with the average deviation.}
\label{mtotrec}
\begin{tabular}{p{3cm}r@{.}lr@{.}l}
Model:				&
\multicolumn{2}{l}{input $M_\tot$}&
\multicolumn{2}{l}{output $M_\tot$} \\
				&
\multicolumn{2}{l}{ $[\log_{10} \Msun]$}&
\multicolumn{2}{l}{$[\log_{10} \Msun]$} \\ \hline
{\bf Fig.~\ref{sfrsigmarel} a)} Constant SFR\\
\quad\quad\quad\quad no age error \\
$0.01 \Msun/\yr$	&	 8&00		&	 8&02  ${+0.02 \atop -0.02}$ \rule[-6pt]{0pt}{18pt}\\
$0.1 \Msun/\yr$	&	 9&00  	&	 9&01	${+0.03 \atop -0.03}$ \rule[-6pt]{0pt}{18pt}\\
$1 \Msun/\yr$		&	10&00 	&	10&01	${+0.04 \atop -0.04}$ \rule[-6pt]{0pt}{18pt}\\
$10 \Msun/\yr$		&	11&00		&	11&01	${+0.05 \atop -0.04}$ \rule[-6pt]{0pt}{18pt}\\ 
{\bf Fig.~\ref{sfrsigmarel} b)}  Three bursts\\
\quad\quad\quad\quad $\sigma_\tau = 0.15\ \dex$ 
					&	9&39		&	 9&40 ${+0.05 \atop -0.04}$ \rule[-6pt]{0pt}{18pt}\\
{\bf Fig.~\ref{sfrsigmarel} c)}  Three bursts\\
\quad\quad\quad\quad $\sigma_\tau = 0.35\ \dex$ 					&	9&39		&	 9&41 ${+0.06 \atop -0.04}$ \rule[-6pt]{0pt}{18pt}\\
{\bf Fig.~\ref{sfrsigmarel} d)}  One burst / Gyr\\
\quad\quad\quad\quad $\sigma_\tau = 0.15\ \dex$					&	10&45		&	10&48${-0.07 \atop -0.05}$ \rule[-6pt]{0pt}{18pt}\\
\end{tabular}
\end{table}

Panels b) and c) of Fig.~\ref{figsfrsigma} show the results for the SFH with three bursts and different age uncertainties.
For the smaller age uncertainty, $\sigma_\tau = 0.15$ dex, all three epochs of enhanced star formation can be identified, i.e. the variation of $\bar{\SFR} (t)$ is comparable or larger than $\sigma_{\SFR\pm}(t)$.
However, the reconstructed shape of the second and third burst is much wider and less pronounced. 
The larger age uncertainties ($\sigma_\tau = 0.35$ dex) lead to a reconstructed SFH where the SFR decreases with age and only allow the reconstruction of the youngest burst.
Due to the age uncertainties $\bar{\SFR} (t)$ drops at old ages, as discussed in Section~\ref{secsfrvar}.
The average deviations $\sigma_{\SFR\pm}$ behave similarly to the constant case.

In the case of one burst every Gyr, lasting for 200 Myr, (Fig.~\ref{figsfrsigma}, panel d) with the small age uncertainty ($\sigma_\tau = 0.15$ dex) only the youngest burst can be detected. 
Then the reconstructed SFH declines until a minimum between the first and second burst is reached.
From the second burst on only a constant SFR with an intermediate value can be recovered.
$\sigma_{\SFR\pm}$ has the same features as before.

A comparison of the averaged recovered SFHs and the input SFHs of these models show the capabilities of our method.
Due to the averaging process structures in the SFH on time scales shorter than the averaging window cannot be reconstructed.
In the other cases the sensitivity for structures in the SFH depends on the quality of the age determination.
For the situation of a SFH only slowly varying or with bursts that are well separated and small age errors, the shape of the underlying SFH can be extracted with our method.
Large age errors of the star clusters or highly variable SFHs do not allow us to recover all of the initial features in the derived SFH.
However, even in these cases the absolute value of the recovered SFR is of the same order of magnitude as the actual one.

Thus, the total recovered stellar mass ($M_{\tot,\mathrm{rec}} = \int \SFR(t) d t$) corresponds to the underlying SFH, summarised in Table~\ref{mtotrec}.
The ensemble average of $M_{\tot,\mathrm{rec}}$ (average of the 1000 galaxies) equals to the input value and no bias is introduced.

\begin{figure}
\includegraphics[width=8cm]{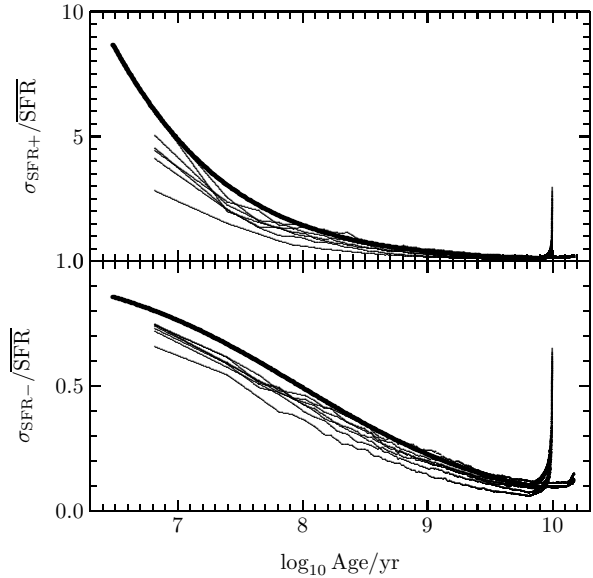}
\caption{Relative errors (thin solid lines) for different constant SFHs with a SFR of 0.01$\Msun/\yr$, 0.1$\Msun/\yr$, 1$\Msun/\yr$ and 10$\Msun/\yr$  and for the bursting cases as above, Fig.~\ref{figsfrsigma}.
The thick solid line shows an upper envelope of the relative error as described in the text (eqs.~\ref{errorfitup} and \ref{errorfitlow}).
}
\label{sfrsigmarel}
\end{figure}

The second aim of the models is to investigate the scatter in the reconstructed SFHs.
Two properties of the average deviations, $\sigma_{\SFR+}$ and $\sigma_{\SFR-}$, are noticeable:
First, in the log-log diagram the positive and negative deviation have for a certain age approximately the same distance to $\bar{\SFR}(t)$.
A reason for this effect could lie in the asymmetric distribution of the most massive star cluster  (Fig.~\ref{figmmaxpdf}).
Thus, using the average of $\sigma_{\SFR+}$ and $\sigma_{\SFR-}$ would lead to a wrong uncertainty estimate.
The second property of both average deviations is that their values relative to $\bar{\SFR}$ are independent of the value of $\bar{\SFR}$.
This is visible in Fig.~\ref{sfrsigmarel}, where the quantities 
\< \hat{\sigma}_{\SFR\pm} (t) &=& \frac{\sigma_{\SFR\pm} (t)}{\bar{\SFR}(t) }, \>
the relative average deviations, are plotted.
To show the independence of $\hat{\sigma}_{\SFR\pm}$ from the shape of a SFH the SFHs from Fig.~\ref{figsfrsigma}, panels b) -- d), are used.
The model with a constant SFR of $0.1 \Msun/\yr$ (Fig.~\ref{figsfrsigma}, panel a)), and additional models with $0.01 \Msun/\yr$, $1 \Msun/\yr$, and $10 \Msun/\yr$ demonstrate the invariance from the absolute value of the SFR.
The independence of the relative average deviations from the shape and absolute value of the SFH makes this quantity  suitable for estimating the uncertainties of the method.
For the implementation in our method for deriving SFHs of individual galaxies we used the analytic fitting formulae
\< \hat{\sigma}_{\SFR+} (\tau) &=& \frac{45}{1 + \exp {1.3(\log_{10} \tau - 5.4) } } \label{errorfitup} \>
and
\< \hat{\sigma}_{\SFR-} (\tau) &=& \frac{1}{1 + \exp{1.2(\log_{10} \tau - 8.0) }}. \label{errorfitlow} \>
These fits are a conservative estimate and lie slightly above the experimental data.
In the method the absolute average deviation  is then calculated by
\< \sigma_{\SFR\pm} (\tau) &=& \hat{\sigma}_{\SFR\pm} (\tau) \times \SFR (\tau). \>
It will be used in the criterion to detect significant variations in a SFH, discussed next.

\subsection{\label{sigcrit} A criterion to detect significant variations in a SFH}

\begin{figure}
\includegraphics[width=8cm]{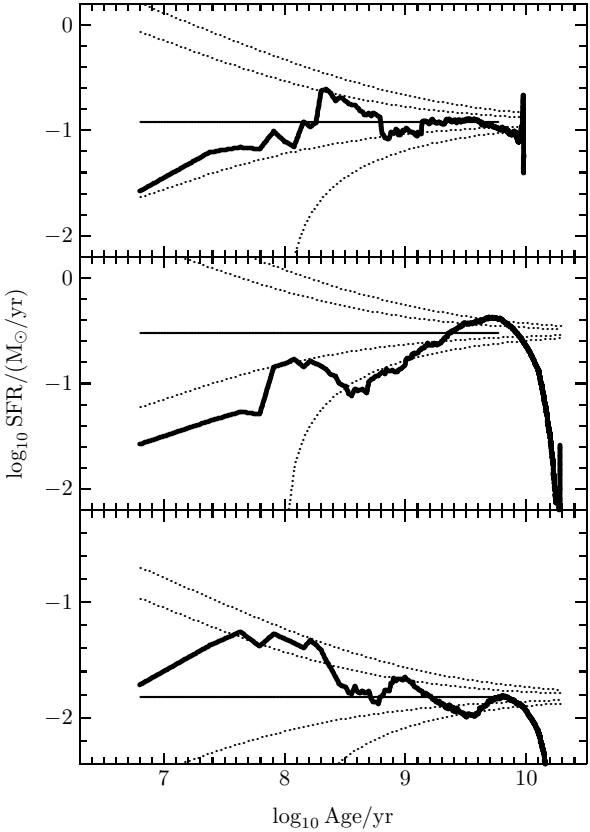}
\caption{
Reconstructed SFHs from Figs.~\ref{figsfrconst}, \ref{figsfrlin}, and \ref{figsfrbump} (top to bottom: constant, linearly decreasing, with three bursts; thick lines), shown with the null hypothesis of a constant SFH (thin solid lines, $\langle \SFR \rangle$).
For a significant variation of the reconstructed SFH it has to cross the $1\sigma_{\SFR\pm}$ line (inner dotted lines).
The outer dotted lines are at $2\sigma_{\SFR\pm}$.
}
\label{figsfhpanelref}
\end{figure}

Due to the probabilistic distribution of the most massive cluster scatter in the reconstructed SFH is expected.
As visible in Figs.~\ref{figsfrconst} and \ref{figsfrlin} the reconstructed SFH can mimic periods of reduced or enhanced star formation.
However, the variations generated in this way are not caused by real events.
The average deviations derived in the previous Section can be used to give the region of SFHs compatible with the reconstructed one. 
But to verify variations in a SFH we suggest to disproof the hypothesis that there are {\it no} variations.
This is done by setting up the null hypothesis of a constant SFH with a SFR equal to the time-averaged SFR, $\langle \SFR \rangle$, of the reconstructed SFH.
If the reconstructed SFH leaves significantly the $1\sigma_{\SFR\pm}$-region of the null hypothesis, a constant SFH, can hardly be supported.
Thus the actual SFH of the galaxy has to have variations.
This is our suggested criterion of significance.
Note that this criterion implies that troughs and maxima that result from a truly variable SFR differ by more than $2 \sigma_{\SFR\pm}$.

$\langle \SFR \rangle$ is given by the integral over the SFH, divided by the time.
With our method two  SFHs are reconstructed, the lower limit with gaps where no cluster was observed, and the upper limit with the fading limit mass used in the gaps.
Therefore $\langle \SFR \rangle$ has to be calculated as the mean of the average SFR of both limits.
In reality these limits start to differ for older ages, caused by incompleteness due to cluster evolution and observational limits. 
Especially in galaxies with strong cluster evolution the upper limiting SFH can shift $\langle \SFR \rangle$ towards unreasonable values if it is integrated over all ages.
To prevent this we integrate only up to the age where the logarithms of the upper and lower limit differ by less than 0.2.

The outcome of this procedure is displayed in Figs.~\ref{figsfhpanelref} and \ref{sfhlmc}.
Figure~\ref{figsfhpanelref} shows again the  SFHs of Figs.~\ref{figsfrconst}, \ref{figsfrlin}, and \ref{figsfrbump}, now with the criterion for variations.
Since in these models no cluster evolution was incorporated, we integrated up to an age of 6 Gyr (where the thin solid line stops) to obtain $\langle \SFR \rangle$.
In the constant case the input SFR ($0.1 \Msun /\yr $) and the average ($\langle \SFR \rangle = 0.12 \Msun / \yr$) are in good agreement.
The extremes of recovered SFH barely exceeds the $1\sigma_{\SFR\pm}$ region.
Thus the null hypothesis of a constant SFH cannot be rejected, as is correct in this case.

The reconstructed linearly decreasing SFH lies, during most times, far away from the average.
At young ages the recovered SFH lies sometimes even below $2 \sigma_{\SFR-}$, and before the decline due to the age uncertainties at old ages it rises above $2 \sigma_{\SFR+}$. 
From this can be deduced that the null hypothesis is not consistent with the reconstructed SFH.
A clear trend of increase with age is visible.

The bursting case is harder to identify.
Only the first burst (0--200 Myr), and the last dip (1.2--8.8 Gyr) leave unambiguous traces, albeit the shape is less pronounced than that of the underlying SFH.
The two other bursts and the first dip merely allow an ``educated guess'' of the shape of the actual SFH.
As above a constant SFH is not consistent with the reconstructed SFH.
However, the {\it exact} structure of this SFH cannot be determined with sufficient certainty, although it is visible.

The above cases show that our suggested criterion of significance allows us to distinguish between features in the actual SFH and artefacts due to the method.

\subsection{\label{lmcmodel} Constant SFR including cluster evolution and an observational limit}

\begin{figure}
\includegraphics[width=8cm]{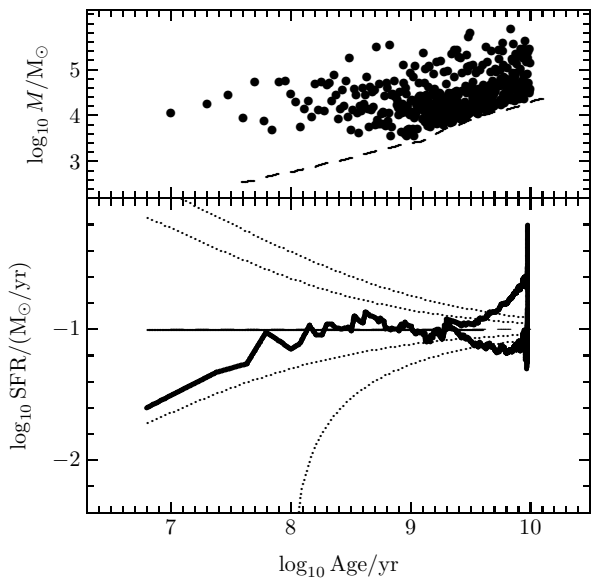}
\caption{Recovered SFH for a LMC-type cluster model that includes dynamical evolution and the missing of clusters due to the observational flux limit (dashed line in the upper part).
As previously the upper panel shows the cluster distribution, whereby here each cluster is shown with its current mass after evolving it dynamically, and the lower panel contains the reconstructed SFH (thick lines).
Due to missing clusters the reconstructed SFH branches in two parts, as described in Section \ref{lmcmodel}.
The thin solid line is the constant SFH ($\langle \SFR \rangle$)used to detect significant variations of the true SFH and stops when the averaging is stopped.
Dotted lines indicate the  $1\sigma_{\SFR\pm}$ and $2\sigma_{\SFR\pm}$ regions.
}
\label{sfhlmc}
\end{figure}

The models described above do not account for cluster evolution and the observational limiting magnitude for cluster detection.
To show the consequences of these effects a galaxy-model with conditions similar to the Large Magellanic Cloud (LMC) was generated.
The stellar mass of the LMC was determined by \citet{kim-etal1998} to be $2.0 \times 10^9 \Msun$.
Assuming a constant SFH leads to a SFR of $\approx 0.1 \Msun/\yr$, which we used in our model.
The cluster-disruption parameter is $t_4 = 7.9 \times 10^9 \yr$ \citep{boutloukos+lamers2003} and the flux limit is $\mathcal{M}_\mathrm{lim} = - 3.5 \mathrm{mag}$ \citep{hunter-etal2003} in the V Band.
The generated data and the results are shown in Fig.~\ref{sfhlmc}.

Because of the weak tidal field clusters evolve only slowly and the high mass-end of the clusters is similar to the case without cluster evolution.
The effect of the observational limit is clearly visible as a cut-off in the lower part of the cluster distribution.
The minimum observable mass increases with time due to the luminosity evolution of the clusters.

As in the previous models the reconstructed SFH deviates from the input value at young ages  because of the small number of formation epochs used for averaging.
Then there is a period of good agreement, until some formation epochs contain no clusters any more.
Due to the cluster evolution the clusters of these epochs are dissolved or have lost such a large fraction of their stellar content that they cannot be detected.
Therefore the reconstructed SFH now shows two branches corresponding to the upper and lower limit.
Assuming that no detection of a cluster means that there was no star formation activity leads to the lower limit.
Using the detection limit as the mass of the most massive cluster gives the upper limit.
The true SFH lies between both limits, which is indeed confirmed:
Until the branching of the upper and lower limit the $1\sigma_{\SFR\pm}$-region contains the reconstructed SFH.
From this point on only a rough estimate of the SFH can be obtained.
From the input SFH a total stellar mass of $1 \times 10^9 \Msun$ was  built up.
The upper and lower limit reconstructed $0.78 \times 10^9 \Msun$ and $1.74 \times 10^9 \Msun$, embracing the model value. 
That the reconstructed mass range contains the known input value constitutes a consistency check.

\section{Conclusions and Summary}
Based on the assumption that all stars form in star clusters it is possible to explain the relation between the brightest clusters in a galaxy and the present star formation rate, assuming that the brightest cluster is also the most massive.
During a formation epoch lasting for $\approx$ 10 Myr a complete population of star clusters is formed.
The mass of the most massive cluster obeys a distribution function that can account for the scatter in the $M_\mx(\SFR)$ relation.
The SFH of a galaxy can then be seen as a sequence of formation epochs.
Starting from this we presented a new method to derive SFHs using star clusters taking into account the statistical properties of the most massive cluster as well as their dynamical evolution.

The method was tested for a number of model SFHs for which synthetic cluster populations were created.
The tests show that our method is capable of reproducing the modelled SFHs if they are only slowly varying or have bursts which are well separated.
To be resolved, the time between two short-time bursts needs to be longer than 0.5 dex, the time over which is averaged.
However, the typical uncertainties in the age determination and the need for averaging do not allow a shorter averaging window.
Artefacts result from averaging over too few small age bins and from missing data at high ages.
The example SFHs show to which degree our method can be used to make confident statements about the SFH of a galaxy.

A model including realistic conditions for observation and cluster evolution also leads to  good agreement between the input and the reconstructed SFH.
In following contributions we will apply this method to the galaxies LMC, SMC, M51 and M101.
For the LMC we will compare this new method to the results obtained using the CMD method.

\section*{Acknowledgements}
We thank Carsten Weidner for useful discussions, and Ylva Schuberth for critical reading of the manuscript.
ThM acknowledges financial support by the AIfA.

\bibliographystyle{mn2e}
\bibliography{../bib/cluster}

\bsp

\label{lastpage}

\end{document}